\documentclass[12pt,a4paper]{article}
\usepackage{amssymb,amsmath}
\usepackage[dvips]{lscape,graphicx}
\usepackage{cite}  
\usepackage{epsfig}
\newcommand{\bc}{\begin{center}}
\newcommand{\ec}{\end{center}}
\newcommand{\bd}{\begin{displaymath}}
\newcommand{\ed}{\end{displaymath}}
\newcommand{\be}{\begin{equation}}
\newcommand{\ee}{\end{equation}}
\newcommand{\lb}{\label}
\newcommand{\ba}{\begin{array}}
\newcommand{\ea}{\end{array}}
\newcommand{\bt}{\begin{tabular}}
\newcommand{\et}{\end{tabular}}
\newcommand{\un}{\underline}
\newcommand{\ov}{\overline}

\begin{document}

\title{\bf Flipped SU(5), see--saw scale physics and degenerate vacua}

\author{C.R.~Das\,${}^{1}$\,\footnote{\small\, crdas@imsc.res.in} ,
C.D.~Froggatt\,${}^{2,\, 4}$\,\footnote{\small\,
c.froggatt@physics.gla.ac.uk} , L.V.~Laperashvili\,${}^{1,\, 3}
$\,\footnote{\small\, laper@imsc.res.in, laper@itep.ru} \, and
H.B.~Nielsen\,${}^{4}$\,\footnote{\small\, hbech@nbi.dk}\\[10mm]
\itshape{${}^{1}$ The Institute of Mathematical Sciences, Chennai,
India}\\[3mm] \itshape{${}^{2}$ Department of
Physics and Astronomy,}\\ \itshape{Glasgow University,
Glasgow, Scotland}\\[3mm] \itshape{${}^{3}$ Institute of
Theoretical and Experimental Physics,
Moscow, Russia}\\[3mm] \itshape{${}^{4}$ The Niels Bohr
Institute, Copenhagen, Denmark }}

\date{}

\maketitle

\begin{abstract}
\noindent

We investigate the requirement of the existence of two degenerate
vacua of the effective potential as a function of the
Weinberg--Salam Higgs scalar field norm, as suggested by the
multiple point principle, in an extension of the Standard Model
including see--saw scale physics. Results are presented from an
investigation of an extension of the Standard Model to the gauge
symmetry group $SU(3)_C\times SU(2)_L\times U(1)'\times \tilde
U(1)$, where two groups $U(1)'$ and $\tilde U(1)$ originate at the
see--saw scale $M_{SS}$, when heavy (right--handed) neutrinos
appear. The consequent unification of the group $SU(3)_C\times
SU(2)_L\times U(1)'$ into the flipped $SU(5)$ at the GUT scale
leads to the group $SU(5)\times \tilde U(1)$. We assume the
position of the second minimum of the effective potential
coincides with the fundamental scale, here taken to be the GUT
scale. We solve the renormalization group equations in the
one--loop approximation and obtain a top--quark mass of $171\pm 3$
GeV and a Higgs mass of $129\pm 4$ GeV, in the case when the
Yukawa couplings of the neutrinos are less than half that of the
top quark at the GUT scale.

\end{abstract}

\clearpage\newpage

For some time \cite{1,2} we have sought to derive the values of
Standard Model (SM) parameters from what we call the Multiple
Point Principle (MPP), according to which there are several vacua
all having exceedingly small cosmological constants like the
vacuum in which we live now. But we have no guarantee for how far
the SM will work up the energy scale. To investigate the influence
of new physics -- especially the see--saw neutrino mass producing
physics \cite{see--saw} -- at higher scales on our predictions 
from MPP, such as
the top--quark mass or the Higgs mass, we shall here investigate a
non--supersymmetric flipped $SU(5)$ extension of the SM broken 
though not in the
normally suggested way by a decuplet (or in SUSY versions even a
couple of conjugate decuplets). Rather we let the flipped $SU(5)$,
which is $SU(5)\times \tilde U(1)$, break stepwise, firstly
highest in energy scale at the unifying scale $M_{GUT}$ by an
adjoint Higgs field down to $SU(3)_C\times SU(2)_L\times
U(1)'\times \tilde U(1)$ and next at a lower see--saw scale
$M_{SS}$ down to the SM group $SU(3)_C \times SU(2)_L\times
U_Y(1)$, say by a Higgs field $S$ which belongs to a $\un{50}$
representation of $SU(5)$. It is our philosophy not to take the details
too seriously but rather think of the flipped $SU(5)$ as a typical
representative model with new physics, which we can use to
estimate the magnitude of the deviations caused to the
MPP--predictions.

The specific stepwise breaking used in the present article is
taken in order to have a see--saw scale as a separate scale which
can be varied. In this respect we do not use the advertised
benefits of the usual flipped $SU(5)$ \cite{see--saw,3,4}, which does not
use an adjoint Higgs field but rather a decuplet Higgs as
favoured by superstring theory.

In flipped $SU(5)$, the quarks and leptons are in the $\un{1},
\un{\bar{5}}, \un{10}$ representations, but with assignments
and electric charges
`flipped' relative to conventional $SU(5)$. In either standard or
`flipped' $SU(5)$ \cite{5,6,7} a single generation of 16 matter fields,
including a singlet right--handed neutrino, can be accommodated by
a set of $\un{1}, \un{\bar{5}}, \un{10}$ representations.
However, the difference
between the flipped and conventional versions of $SU(5)$ is in the
way in which the 16 matter fields of each generation are embedded
in these representations. Flipped $SU(5)$ received its name from the
exchanges  in the assignments of the fields: up--like and
down--like fields are exchanged, as are electron--like with
neutrino--like, as well as their anti--particle partners.
The particle content of the flipped $SU(5)\times {\tilde U}(1)$
model used here is as follows
\begin{itemize}
\item[1.] three generations of matter fields:
\be
    F = \left(\un{10},\, \frac{1}{2\sqrt{10}}\right), \qquad 
{\bar f}_i = \left(\un{\bar 5},
\, -\frac{3}{2\sqrt{10}}\right), \qquad l_i^c = \left(\un{1},\, \frac
{5}{2\sqrt{10}}\right), \qquad (i=1,2,3), \lb{1} \ee
\item[2.] a five--dimensional (Weinberg--Salam) Higgs to break
$SU(2)_L\times U(1)_Y$: 
\be        
\phi = \left(\un{5},\,-\frac{1}{\sqrt{10}}\right),  \lb{2} 
\ee
\item[3.] an adjoint 24--dimensional Higgs to break $SU(5)\times \tilde
U(1) \to SU(3)_C\times SU(2)_L\times U(1)'\times \tilde U(1):$ 
\be
A= (\un{24},\,0) \lb{3}, \ee
\item[4.] a Higgs field $S$ to break $SU(3)_C\times SU(2)_L\times
U(1)'\times \tilde U(1) \to  SU(3)_C\times SU(2)_L\times U(1)_Y$
at the see--saw scale, with the following $U(1)'\times \tilde U(1)$
quantum numbers:
\be Q'_S = 2\sqrt{\frac{3}{5}}, \qquad \tilde Q_S
= \sqrt{\frac{1}{10}}.  \lb{4}
\ee 
\end{itemize}
We note that we have chosen to
normalise all the flipped $SU(5)$ generators $ T_a$ such that the
trace of $T_a^2$ over the 16 fermions in a single quark--lepton
generation is given by $Tr_{16}(T_a^2)=2$.

We do not attempt here to solve the fermion mass problem, which
would need extra new physics \cite{fn} at, say, the GUT scale.
However phenomenologically we know that, apart from the top quark,
the Yukawa couplings of charged fermions can be neglected.
Unfortunately, we do not have direct information about the Yukawa
couplings of the neutrinos. In a naive minimal flipped $SU(5)$ model
one might expect that the Dirac neutrino mass matrix would be
equal to the up quark mass matrix. But the see--saw mechanism would
then almost inevitably give an unrealistically strong hierarchy of
light neutrino masses. Since we cannot extract reliable values for
the neutrino Yukawa couplings, we allow the possibility that one
of them $y_{\nu}$ might be as large as the top quark Yukawa
coupling $h$ and introduce the parameter $p$ giving their ratio at
the GUT scale $M_{GUT}$: \be y_{\nu}(M_{GUT}) = p \cdot h(M_{GUT}),
\lb{p} \ee where $0 \le p \le 1$.

The renormalization group equations (RGEs) are: \be
       \frac{dg_i}{dt} = \beta_{g_i}, \qquad
       \frac{dh}{dt} = \beta_h, \qquad 
       \frac{d\lambda}{dt} = \beta_{\lambda}, \qquad
       \frac{dy_{\nu}}{dt} = \beta_{y_{\nu}},
         \lb{8}
\ee where $t=\ln(\mu/M)=\ln(\phi/M)$ is the evolution variable,
$\mu$ is the energy scale, $M$ is the renormalization mass scale
and $\phi$ is the Weinberg--Salam Higgs scalar field. Its vacuum
expectation value (VEV) is: \be
              <\phi> = \frac{1}{\sqrt 2}\left(
             \ba{c}
             0\\
             {\it v}
             \ea
             \right),   \qquad {\mbox{with}} \qquad v\approx
             246 \ \mbox{GeV}.\lb{9}
\ee The gauge couplings $g_i=(g_1,g_2,g_3)$ correspond to the
$U(1)$, $SU(2)_L$ and $SU(3)_C$ groups of the SM, and $\lambda$
is the Weinberg--Salam Higgs field self--interaction coupling
constant. We neglect the Yukawa couplings of all the other
fermions. Also we neglect interactions of the form $\phi^2S^2$
between the Higgs fields.

In the Weinberg--Salam theory the tree level masses of the gauge
bosons $W$ and $Z$, the top quark and the physical Higgs boson $H$
are expressed in terms of the VEV parameter $v$: 
\be
          M_W^2 = \frac{1}{4} g^2 v^2,
 \qquad
          M_Z^2 = \frac{1}{4} \left(g^2 + g'^2\right) v^2,
 \qquad
          m_t = \frac{1}{\sqrt 2} h v,
 \qquad
          m_H^2 = \lambda v^2,         \lb{10}
\ee where $g'\equiv g_Y = \sqrt {(3/5)}\,g_1$, $g\equiv g_2$.
The one--loop $\beta$--functions in the SM are: \be 16\pi^2
\beta_{g_1}^{(1)} = \frac{41}{10}g_1^3, \qquad 16\pi^2
\beta_{g_2}^{(1)} = - \frac{19}{6}g_2^3, \qquad 16\pi^2
\beta_{g_3}^{(1)} = - 7 g_3^3, \lb{11} \ee \be 16\pi^2
\beta_{h}^{(1)} = h\left(\frac{9}{2} h^2 - 8g_3^2 - \frac{9}{4} g_2^2 -
\frac{17}{20} g_1^2\right), \lb{13} \ee \be 16\pi^2\beta_{\lambda}^{(1)}
= 12\lambda^2 + \lambda \left(12h^2 - 9g_2^2 - \frac{9}{5}g_1^2\right) +
\frac{27}{100}g_1^4 + \frac{9}{10}g_1^2 g_2^2 + \frac{9}{4}g_2^4 -
12h^4. \lb{12} \ee These equations are valid up to the see--saw
scale $M_{SS}$.

In the region from $M_{SS}$ to $M_{GUT}$ we have a new type of
symmetry $SU(3)_C\times SU(2)_L\times U(1)'\times \tilde U(1)$
with the following one--loop $\beta$--functions for the
corresponding RGEs similar to (\ref{8}): \be 16\pi^2\beta_
{{g'}_1}^{(1)} = \frac{45}{10}{g'}_1^3, \qquad 16\pi^2
\beta_{\tilde g_1}^{(1)} = \frac{41}{10}{\tilde g_1}^3, \qquad
 16\pi^2\beta_{g_2}^{(1)} = - \frac{19}{6}g_2^3, \qquad 16\pi^2
\beta_{g_3}^{(1)} = - 7 g_3^3, \lb{14} \ee \be 16\pi^2
\beta_{h}^{(1)} = h\left(\frac{9}{2} h^2 - 8g_3^2 - \frac{9}{4} g_2^2 -
\frac{1}{4} {g'}_1^2 - \frac{3}{4}{\tilde g}_1^2\right), \lb{16} \ee \be
16\pi^2 \beta_{y_{\nu}}^{(1)} = y_{\nu}\left(\frac{5}{2} y_{\nu}^2 -
\frac{9}{4} g_2^2 - \frac{3}{4} {g'}_1^2 - \frac{1}{4}{\tilde
g}_1^2\right), \lb{16a} \ee \begin{eqnarray}
16\pi^2\beta_{\lambda}^{(1)}&=& 12\lambda^2 + \lambda \left[12h^2 +
4y_{\nu}^2 - 9g_2^2 - \frac{9}{5}\left({g'}_1^2 + \frac{2}{3}{\tilde
g}_1^2\right)\right] + 
\frac{27}{100}\left({g'}_1^2 + \frac{2}{3}{\tilde g}_ 1^2\right)^2
\nonumber
\\ && + \frac{9}{10}\left({g'}_1^2 + \frac{2}{3}{\tilde g}_ 1^2\right)g_2^2+
\frac{9}{4}g_2^4 - 12h^4-4y_{\nu}^4. \lb{15} \end{eqnarray}  Here
we neglected small couplings.

Following the idea of Refs.~\cite{1,2}, we assume that in the present
model the fundamental scale $M_{fund}$ coincides with the GUT
scale $M_{GUT}$ for $SU(5)\times {\tilde U}(1)$.
This idea is based on the Multiple Point
Principle (MPP) \cite{8} (see also the reviews \cite{9,9a}), 
according to which
several vacuum states with the same energy density exist in
Nature. In the pure SM the effective potential for the
Weinberg--Salam Higgs field can have two degenerate minima as a
function of $|\phi|$: \be
        V_{eff}(\phi_{min1}^2) = V_{eff}(\phi_{min2}^2) = 0,       \lb{17}
\ee \be
        V'_{eff}(\phi_{min1}^2) = V'_{eff}(\phi_{min2}^2) = 0,       \lb{18}
\ee where \be
         V'(\phi^2) = \frac{\partial V}{\partial \phi^2}.
                                             \lb{19}
\ee The first minimum is the standard ``Weak scale minimum", and
the second one is the non--standard ``Fundamental scale minimum" as
shown in Fig.~\ref{fig1}. In the present model the assumption is that the
second minimum of the effective potential coincides with the
GUT--scale $\phi_{min2} = M_{GUT}$.

As discussed in Ref.~\cite{1}, for large values of the Higgs field
$\phi^2 \gg v^2$ the degeneracy conditions (\ref{17}) and
(\ref{18}) lead to the following requirements: \be
       \lambda(\phi_{min2}) = 0, \qquad
        \beta_{\lambda}(\phi_{min2}, \lambda=0) = 0.  \lb{20}
\ee Taking $\phi_{min2} \sim M_{Planck}$ and using the  two--loop
RGEs, the following MPP predictions were obtained in the pure SM
for the top quark and Weinberg--Salam Higgs particle masses
\cite{1,10}: 
\be M_t = 173 \pm 5 \ \mbox{GeV}, \qquad M_H = 135
\pm 9 \ \mbox{GeV}. \lb{fn} \ee We note that the present
experimental value \cite{cdf} of the top quark mass is: 
\be 
M_t = 172.7 \pm 2.7.   \lb{mt} \ee
When solving the RGEs (\ref{14}--\ref{15}) we use the MPP
conditions (\ref{20}) at the GUT scale, which in our case
determine the top quark Yukawa coupling at the GUT scale in terms
of the $SU(2)_L\times U(1)'\times \tilde U(1)$ gauge couplings at
the GUT scale and the parameter $p$: 
\be
4(3+p^4)h^4=\frac{27}{100}\left({g'}_1^2 + \frac{2}{3}{\tilde g}_
1^2\right)^2 + \frac{9}{10}\left({g'}_1^2 + \frac{2}{3}{\tilde g}_ 1^2
\right)g_2^2+ \frac{9}{4}g_2^4. \lb{hgut} 
\ee By considering the
joint solution of
the RGEs (\ref{11}--\ref{15}), we estimate corrections to the MPP
predictions for $M_t$ and $M_H$, due to the new see--saw scale
physics.

Starting from the Particle Data Group \cite{11}, the
phenomenological input to our calculations are the $Z^0$ mass: \be
      M_Z = 91.1876 \pm 0.0021\ {\mbox{GeV}}, \lb{21}
\ee the inverse electromagnetic fine structure constant and the
square of the sine of the weak angle in the $\ov{\rm {MS}}$--scheme:
\be
         {\hat \alpha}^{-1}(M_Z) = 127.906 \pm 0.019, \qquad 
    {\hat s}^2(M_Z) = 0.23120 \pm 0.00015,   \lb{22}
\ee and the QCD fine structure constant: \be
       \alpha_3(M_Z) = 0.119 \pm 0.002.              \lb{23}
\ee The only other input to our calculation is the value of the
parameter $p$.

It is well--known that the running of all the gauge coupling
constants in the SM is well described by the one--loop
approximation. For $M_t\le \mu \le M_{SS}$ we have the following
evolutions for the inverses of the fine structure constants
$\alpha_i=g_i^2/{4\pi}$, $(i=1,2,3)$, which are revised using
the updated experimental results \cite{11}: \be
      \alpha_1^{-1}(t) = 58.65 \pm 0.02 - \frac{41}{20\pi}t,   
                                           \lb{27}
\ee \be
      \alpha_2^{-1}(t) = 29.95 \pm 0.02 + \frac{19}{12\pi}t,     
                                            \lb{28}
\ee \be
      \alpha_3^{-1}(t) = 9.17 \pm 0.20 + \frac{7}{2\pi}t,
                                           \lb{29}
\ee where $t = \ln(\mu/{M_t}).$ The gauge coupling constant
evolutions (\ref{27}--\ref{29}) are shown in Fig.~\ref{fig2}, where $x =
\log_{10}\mu$ (GeV) and $t=x\ln 10 - \ln {M_t}$. The evolutions
(\ref{28}) and (\ref{29}) are valid up to the GUT scale for $M_t
\le \mu \le M_{GUT}$. But Eq.~(\ref{27}) works only up to the
see--saw scale. At the scale $M_{SS}$ the two $U(1)$ groups
-- $U(1)'$ and $\tilde U(1)$
-- become active and give, instead of Eq.~(\ref{27}), the following new
fine structure constant evolutions:
\be
  {\alpha'}_1^{-1}(t) = {\alpha'}_1^{-1}(M_{SS}) - \frac{45}{20\pi}{\tilde t},
                                           \lb{31}
\ee \be
  {\tilde \alpha}_1^{-1}(t) = {\tilde \alpha}_1^{-1}(M_{SS}) -
  \frac{41}{20\pi}{\tilde t},
                                           \lb{32}
\ee which are valid for the interval $M_{SS} \le \mu \le M_{GUT}$. In
Eq.~(\ref{32}) $$\tilde t =\ln\left(\frac{\mu}{M_{SS}}\right) = t +
\ln\left(\frac{M_t}{M_{SS}}\right)= x\ln 10 - \ln M_{SS}.$$ For convenience
we have also considered the evolution of: \be
y_{top}(t)=\alpha_h^{-1}(t)= \left(\frac{h^2(t)}{4\pi}\right)^{-1}. \lb{36} \ee
The mixture of the two $U(1)$ groups at the see--saw scale leads to
the following relation (compare with Refs.~\cite{3,4}):\be
\alpha_1^{-1}(M_{SS}) = \frac{24}{25}{\tilde
\alpha}_1^{-1}(M_{SS}) + \frac{1}{25}{\alpha'}_1^{-1}(M_{SS}).
                                                           \lb{33}
\ee At the GUT scale we have the unification $SU(3)_C\times
SU(2)_L\times U(1)'\to SU(5)$ giving: \be
  {\alpha'}_1(M_{GUT}) = \alpha_2(M_{GUT}) = \alpha_3(M_{GUT}) =
  \alpha_{GUT}.
                                    \lb{34}
\ee The GUT scale is given by the intersection of the evolutions
(\ref{28}) and (\ref{29}) for $\alpha_2^{-1}(x)$ and
$\alpha_3^{-1}(x)$.

From Eq.~(\ref{33}), using Eqs.~(\ref{31}), (\ref{32}) and relations
(\ref{34}), we obtain the following relation: \be
   \tilde \alpha_1^{-1}(M_{GUT}) =
   \frac{25}{24}\alpha_1^{-1}(M_{GUT}) -
   \frac{1}{24}\alpha_{GUT}^{-1} -
   \frac{1}{120\pi}\ln\left(\frac{M_{GUT}}{M_{SS}}\right).
                                                 \lb{35}
\ee
where the RGE for $\alpha_1(t)$ has been formally extended
to the GUT scale.

We now investigate the dependence of the MPP predictions on the
see--saw scale physics, by varying the see--saw scale $M_{SS}$ and
the parameter $p$ which determines the neutrino Yukawa coupling
$y_{\nu}$. Once $M_{SS}$ and $p$ are fixed, the values of all the
gauge couplings can be determined at the GUT scale, where we also
have the boundary values (\ref{hgut}), (\ref{p}) and (\ref{20})
for $h(M_{GUT})$, $y_{\nu}(M_{GUT})$ and $\lambda(M_{GUT})$
respectively. The RGEs (\ref{11}--\ref{15}) can then be
integrated down from the GUT scale, requiring continuity at the
see--saw scale, to the electroweak scale. In this way we determine
the running top quark mass $m_t(\mu=m_t)=h(\mu=m_t)v/\sqrt{2}$ and
the Higgs self--coupling $\lambda(m_t)$. We can then calculate
\cite{10,12} the top quark pole mass $M_t$ and the Higgs pole mass
$M_H$.

We find that our results are highly insensitive to the value of
the see--saw scale $M_{SS}$, which is allowed to range from 
10 TeV to the GUT scale ($M_{GUT} \sim 10^{17}$ GeV). 
However, as a consequence of Eq.~(\ref{hgut}), there 
is a significant dependence on $p$ for $p \sim 1$. So 
we present below our results for three values of $p$:
\begin{eqnarray}
p=0 \qquad M_t = 171 \pm 3 \ \mbox{GeV} \qquad M_H = 129
\pm4 \ \mbox{GeV}, \\
p=1/2 \qquad M_t = 169 \pm 3 \ \mbox{GeV} \qquad M_H = 128
\pm 4 \ \mbox{GeV}, \\p=1 \qquad M_t = 164 \pm 3 \ \mbox{GeV}
\qquad M_H = 118 \pm 4 \ \mbox{GeV}. \end{eqnarray}
We see that for $p \le 1/2$, the results are essentially 
independent of the new see--saw scale physics. Comparing to the 
pure SM degenerate vacuum based prediction
(\ref{fn}), it means that taking the second minimum (assumed
degenerate with the present vacuum) down from the Planck to the
GUT scale and including the $\beta$--function effects of our flipped
$SU(5)$ only shifts the Higgs mass down by 6 GeV, predicting the
top quark mass of $171\pm 3$ GeV  in agreement with experiment 
\cite{cdf}. However for $p=1$ the MPP prediction for the top 
quark mass is reduced to $164 \pm 3$ GeV, which is disfavoured by 
experiment, although the corresponding Higgs mass of 
$118 \pm 4$ GeV is close to the signal observed by the Aleph 
group at LEP \cite{aleph}. 

Fig.~\ref{fig3} shows an example of the evolutions of $\lambda(x)$ and
$y_{top}(x)$ for $M_t=171$ GeV, $M_H=129$ GeV and
$\alpha_s(M_Z)=0.119$, with $p=0$ and $M_{SS} \sim 10^{11}$ GeV. 
Fig.~\ref{fig2} presents the running of all the fine
structure constants for the same experimental parameters.

We intend to estimate the effects of including two--loop
contributions in the RGEs and to investigate the effect of varying
the position of the second minimum relative to the GUT scale (e.g.
up to the Planck scale).
However it is already clear that the MPP predictions
for the top quark and Higgs masses are rather insensitive to the
introduction of new see--saw
scale physics associated with neutrino masses, unless one of the 
neutrino Yukawa couplings is at least similar in magnitude to 
that of the top quark.

\noindent {\large \bf Acknowledgements:~}
One of the author (LVL) thanks the Institute of Mathematical Sciences,
Chennai (India) for hospitality and financial support. CRD thanks
Prof. G. Rajasekaran and Prof. U. Sarkar for useful discussions.
 This work is supported
by the Russian Foundation for Basic Research (RFBR), project
No.~05--02--17642. CDF would like to acknowledge the hospitality of the
Niels Bohr Institute and support from the Niels Bohr Institute Fund
and PPARC.

\clearpage\newpage
\begin{figure}
\centering
\includegraphics[height=100mm,keepaspectratio=true]{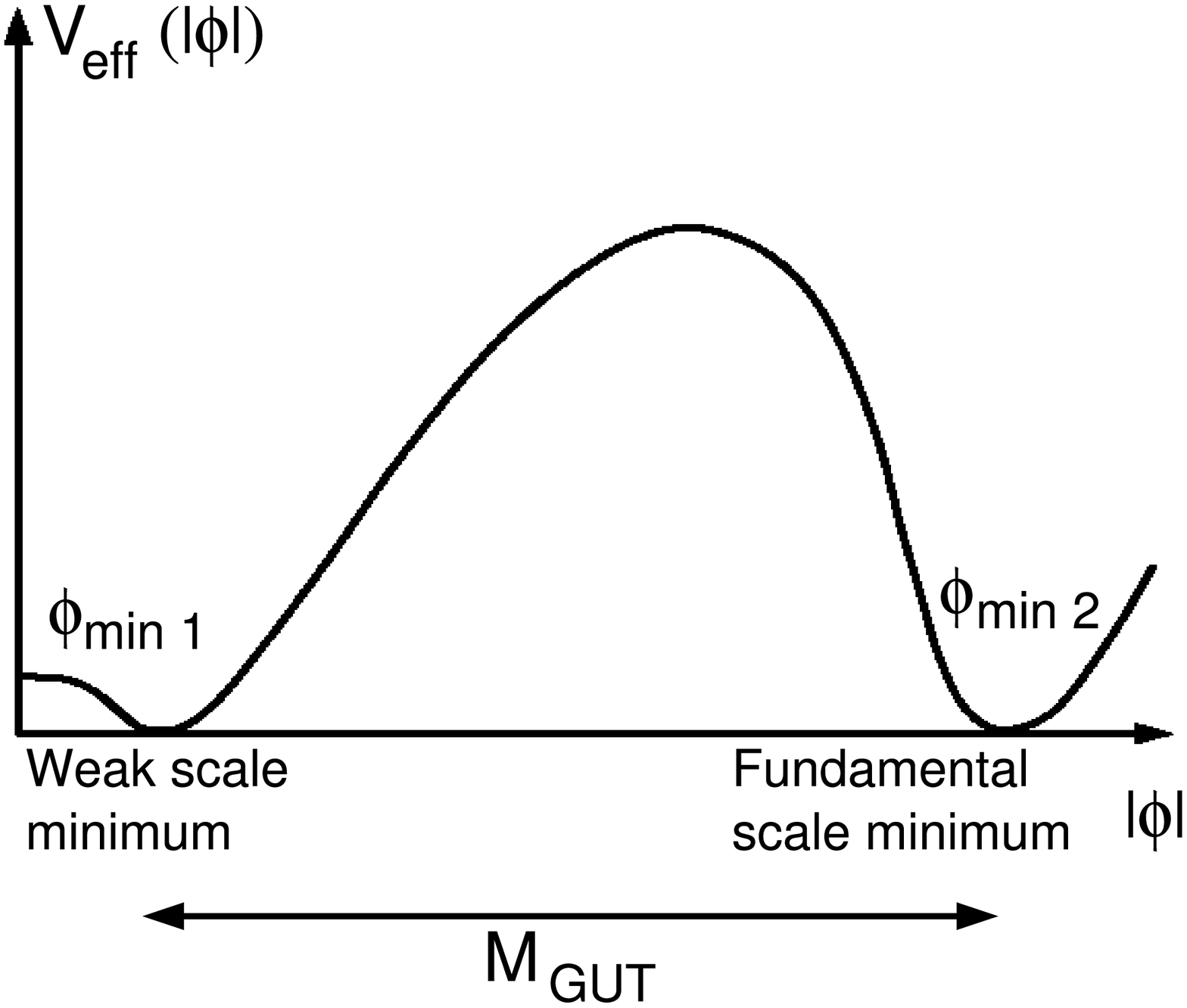}
\caption{Fundamental scale vacuum degenerate with usual SM weak scale vacuum.}
\lb{fig1}
\end{figure}

\clearpage\newpage

\begin{figure}
\centering
{\hspace*{-6mm}
\includegraphics[height=170mm,keepaspectratio=true,angle=-90]{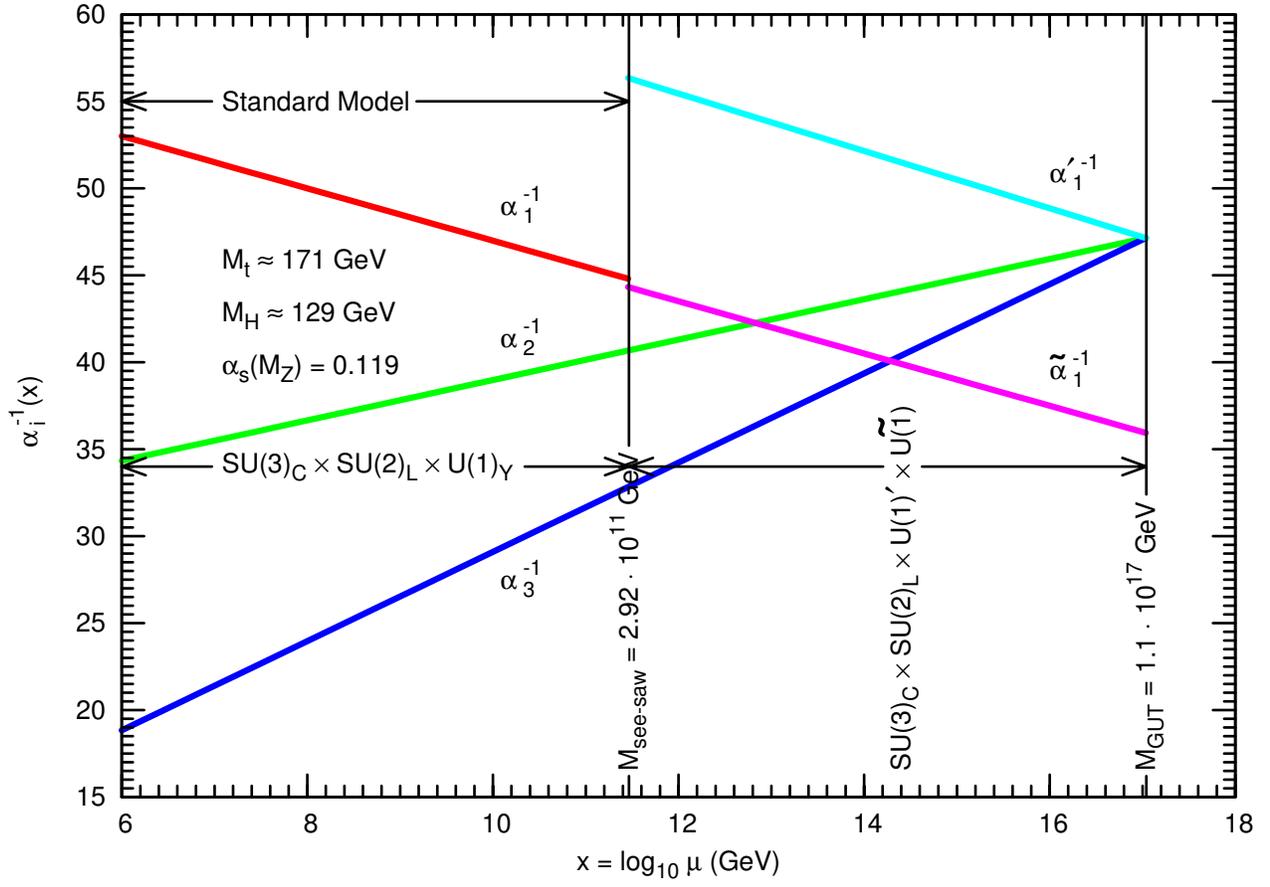}}
\caption{Evolution of the running inverse fine structure constants,
showing the appearance of the $U(1)'\times \tilde U(1)$ gauge
symmetry at the see--saw scale.}
\lb{fig2}
\end{figure}

\clearpage\newpage
\begin{figure}
\centering
\includegraphics[height=235mm,keepaspectratio=true]{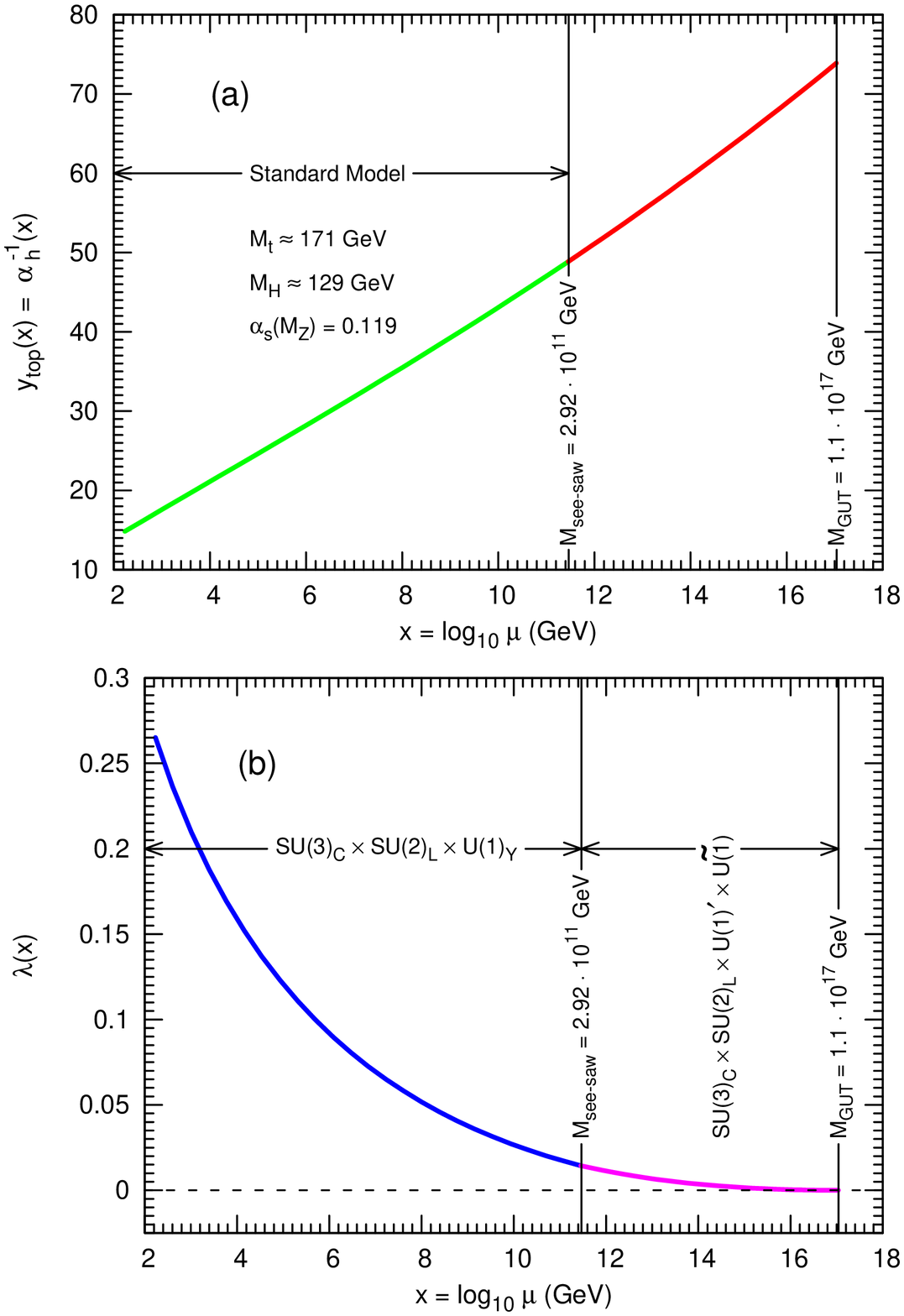}
\caption{Evolution of (a) $y_{top}(x) = \alpha^{-1}_h(x) 
= 4\pi/h^2(x)$ for the top quark and (b) the Weinberg--Salam Higgs
self--coupling constant $\lambda(x)$ for the case $p=0$.}
\lb{fig3}
\end{figure}

\end{document}